# Ultra-high temperature Soret effect in a silicate melt: $SiO_2$ migration to cold side

Yuma Nishida[1], Masahiro Shimizu[1,*], Tatsuya Okuno[1], Jun Matsuoka[2], Yasuhiko Shimotsuma[1], and Kiyotaka Miura[1]

[1] Department of Material Chemistry, Graduate School of Engineering, Kyoto University, Katsura, Nishikyo-ku, Kyoto 615-8510, Japan

[2] School of Engineering, The University of Shiga Prefecture, Hikone 522-8533, Japan

*corresponding author: shimizu.masahiro.3m@kyoto-u.ac.jp

**Abstract**

The Soret effect, temperature gradient driven diffusion, in silicate melts has been investigated intensively in the earth sciences from the 1980s. The $SiO_2$ component is generally concentrated in the hotter region of silicate melts under a temperature gradient. Here, we report that at ultra-high temperatures above approximately 3000 K, $SiO_2$ becomes concentrated in the colder region of the silicate melts under a temperature gradient. The interior of an aluminosilicate glass ($63.3SiO_2$-$16.3Al_2O_3$-$20.4CaO$(mol%)) was irradiated with a 250 kHz femtosecond laser pulse for local heating. $SiO_2$ migrated to the colder region during irradiation with an 800 pulse (3.2 ms irradiation). The temperature analysis indicated that migration to the colder region occurred above 3060 K. In the non-equilibrium molecular dynamics (NEMD) simulation, $SiO_2$ migrated to the colder region under a temperature gradient, which had an average temperature of 4000 K; this result supports the experimental result. On the other hand, $SiO_2$ exhibited a tendency to migrate to the hotter region at around 2400 K in both the NEMD and experimental study. The second-order like phase transition was observed at ~ 2000–3400 K when calculated using MD without a temperature gradient. Therefore, the second-order phase transition could be related to the migration of $SiO_2$ to colder region. However, the detailed mechanism has not been elucidated.

# Ⅰ. INTRODUCTION

The Soret effect is a temperature gradient-driven diffusion that was discovered by Ludwig [1] and tested by Soret [2], and the effect is quantified using the Soret coefficient. The equation for the binary system follows [3]:

$$\sigma_{\text{Soret},1} = -\frac{1}{x_1(1-x_1)} \frac{|\nabla x_1|}{|\nabla T|} \qquad (1)$$

, where, $x$, is the mole fraction of the component, the subscript in $x_1$ indicates the index of the component, | | is the algebraic norm of the vector, and $T$ is the absolute temperature.

Chapman formulated the Soret coefficient of a gas state based on the kinetic theory of gas molecules and concluded that larger and heavier molecules prefer the colder region of the gas under a temperature gradient [4]. Although many models have been developed to predict the Soret coefficient for liquid state [5-18], an accurate model for the liquid state is difficult to accomplish, because of the strong and complex interactions between the diffusion species.

Several reports stemming from the 1980s and since, have indicated that the $SiO_2$ component migrates to the hot region in natural silicate melts [19-23]. For binary silicate melts, two studies on $SiO_2$-$CaO$ [24] and $SiO_2$-$Na_2O$ [25] experimentally demonstrated the migration of $SiO_2$ to the hotter region. The tendency of $SiO_2$ to migrate to the hotter region appeared to be universal. However, molecular dynamics simulations have predicted that $SiO_2$ migrates to the colder regions in specific compositions of $SiO_2$-poor $SiO_2$-$Na_2O$ [25] and $SiO_2$-poor $SiO_2$-$CaO$ [26]. In 2022, Shimizu et al. reported that $SiO_2$ migrates to the colder region in $SiO_2$-poor $SiO_2$-$B_2O_3$ melts [27], and that the migration tendency can be predicted successfully using the modified Kempers model [25,27]:

$$\sigma_{\text{soret},1} = \frac{v_1 v_2}{v_1 x_1 + v_2 x_2} \frac{\frac{h_2 - h_2^\circ}{v_2} - \frac{h_1 - h_1^\circ}{v_1}}{T x_1 \frac{\partial(\mu_1 - \mu_1^\circ)}{\partial x_1}} \quad \ldots (2)$$

, where, $v$ is the partial molar volume, $h$, is the partial molar enthalpy, $\mu$, is the chemical potential, and ° is the symbol indicating the pure liquid state. The modified Kempers model can predict the Soret coefficient of nonideal molecular liquids [28] and several oxide melts such as $SiO_2$-$Na_2O$ [25], $SiO_2$-$CaO$ [26], $B_2O_3$-$Na_2O$ [29], and $GeO_2$-$Na_2O$ [30]. For the silicate melts that exhibit a negative enthalpy of mixing, the condition for the $SiO_2$ migrates to colder region is a "$SiO_2$-poor composition." For the melts comprising $SiO_2$-poor composition, the partial molar enthalpy of $SiO_2$ should be negative, which would lead to a positive Soret coefficient in Eq. (2), and indicates that $SiO_2$ migrates to the colder region.

The significance of $SiO_2$ migration under different temperature gradients in glass engineering is explained next. Once the Soret effect is understood in silicates, one of the factors of compositional inhomogeneity in glass melting tanks and crucibles and damage to laser fabrication in glass may be clarified. Furthermore, the composition of a glass affects many properties of the glass, including the refractive index, chemical durability, phase-separation properties, luminescence, viscosity, heat capacity, elastic properties, and thermal conductivity [31]. From the perspective of applications, laser local heating is an effective method for inducing the Soret effect inside glass. In 2011, Shimizu et al. [24] reported that laser-induced element distribution arises from the Soret effect; Yonesaki et al. [32] discovered laser-induced element redistribution in 2005. There are two advantages of laser local heating: only a short time is needed for the system to achieve steady state (approximately 1 s) because the melting zone is several tens of micrometers in size; and a large mole-fraction change (approximately 10 mol%) is caused by a steep temperature gradient [33]. The fabrication of optical waveguides, utilizing these advantages, has been reported in nonsilicate systems [34,35,36]. Appen's additive factor of the refractive index of $SiO_2$ is smaller than that of the other components [37], which indicates that it is generally difficult to obtain a core-cladding-type optical waveguide using the Soret effect; the difficulty ensues because $SiO_2$ diffuses to the central area of the irradiated zone. $SiO_2$ migration to the colder region is ideal to fabricate the core-cladding type optical waveguides by laser writing.

Fernandez et al. reported that the migration of $SiO_2$ to the surrounding region occurs during high-speed scanning irradiation [38-40]. The complex relationship between temperature and ionic diffusion could not be clarified because high-speed scanning involves moving the heat source. In this study, we fixed the heat source and quantitatively demonstrated $SiO_2$ migration to the colder region at ultra-high temperatures. Fernandez et al. used a composition of $63.3SiO_2$-$16.3Al_2O_3$-$9.7Na_2O$-$10.7CaO$(mol%) [39]. In this study, we simplified the target composition to $63.3SiO_2$-$16.3Al_2O_3$-$20.4CaO$(mol%) to clarify the mechanism.

## Ⅱ. METHOD

### A. Laser irradiation

The $63.3SiO_2$-$16.3Al_2O_3$-$20.4CaO$(mol%) glass was prepared using a melt-quenching method. $SiO_2$, $Al_2O_3$, and $CaCO_3$ powders were mixed and melted in a Pt crucible at 1923 K. The bottom of the crucible was immersed in water for rapid cooling, because the glass melt could not be cast because of the high viscosity of the melts.

After polishing the surface, the glass was irradiated with a Ti:sapphire mode-locked femtosecond laser (Coherent, Mira and RegA) with pulse energy, pulse width, repetition rate, and wavelength of 3.1 μJ, 60 fs, 250 kHz, and 800 nm. The laser was irradiated through a 60× water-immersed objective lens (Nikon, NA = 1.20) at a depth of 170 μm from the surface. The glass is transparent to the wavelength; therefore, when the laser was irradiated with the objective lens, absorption occurred only around the focal spot inside the glass through a multiphoton absorption process. The irradiation time was changed to 3.2 ms, 32 ms, and 1 s, which corresponded to 800, 8000, and 250000 pulses, respectively.

The element distributions of Si, Al, Ca, and O were analyzed using an electron-probe microanalyzer (EPMA, JEOL, JSM-8000), after polishing the glass until the irradiated area was on the surface. Characteristic rays were detected by the wavelength-dispersive X-ray spectroscopy (WDX) method using crystals of TAP, TAP, PETH, and LDE1H for Si, Al, Ca, and O, respectively.

**B. Temperature distribution analysis**

The method for calculating the temperature distribution inside the glass under laser irradiation was similar to that used in previous studies [27,41,42]. The transmission loss analysis was measured with a laser power meter (Coherent, FieldMate)[42]; the absorbed pulse energy of each pulse was 1.8 µJ, and therefore, the absorptivity was 58.6%. The distribution of the thermal energy density, immediately after energy equilibration between the electron and lattice, was assumed to follow the Gaussian function below:

$$\frac{\partial q_{\text{laser}}(t,\boldsymbol{r})}{\partial t} = \delta(t - n\Delta t_L) q_{\text{center}} \exp\left[-\frac{r^2}{\left(\frac{w_r}{2}\right)^2} - \frac{z^2}{\left(\frac{l_z}{2}\right)^2}\right] \quad (3).$$

where, $t$ is the time from the start of the laser irradiation, $q_{\text{laser}}(t,\boldsymbol{r})$ is the thermal energy per unit volume generated by a laser pulse, $\Delta t_L$ is the pulse interval ($= 4$ µs), $q_{\text{center}}$ is the energy density absorbed at the center of the focal spot, $r$ $(= (x^2 + y^2)^{1/2})$ is the radial distance, $z$ is the position along the beam propagation axis, and $w_r$ and $l_z$ are the widths of the absorbed energy in the radial and beam propagation directions, respectively. The values of $w_r$ and $l_z$ were 0.66 and 0.66 µm, respectively. The following equation was used to calculate the energy absorption and thermal conduction:

$$\frac{\partial q(t,\boldsymbol{r})}{\partial t} = \nabla\left(D_{\text{th}} \nabla q(t,\boldsymbol{r})\right) + \frac{\partial q_{\text{laser}}(t,\boldsymbol{r})}{\partial t} \quad (4),$$

, where, q($t$,$\boldsymbol{r}$) is the thermal energy per unit volume and $D_{\text{th}}$ is the thermal diffusivity. The time evolution of the thermal energy density of each pulse was obtained as follows:

$$\Delta q_1(t',\boldsymbol{r}) = q_{\text{center}} \frac{\left(\frac{w_r}{2}\right)^2}{\left(\frac{w_r}{2}\right)^2 + 4D_{\text{th}}t'} \cdot \left[\frac{\left(\frac{l_z}{2}\right)^2}{\left(\frac{l_z}{2}\right)^2 + 4D_{\text{th}}t'}\right]^{\frac{1}{2}} \exp\left[-\frac{r^2}{\left(\frac{w_r}{2}\right)^2 + 4D_{\text{th}}t'} - \frac{z^2}{\left(\frac{l_z}{2}\right)^2 + 4D_{\text{th}}t'}\right] \quad (5)$$

In this study, we used a 250 kHz laser, with a pulse interval of 4 µs. More than 4 µs was required for the thermal energy of one pulse to diffuse out of the surrounding glass matrix because of the low thermal diffusivity of the glass. Under these conditions, the thermal energy of each pulse accumulated, and the total energy density can be expressed as the sum of the contributions of many pulses:

$$q_N(t,\boldsymbol{r}) = \sum_{i=0}^{N-1} \Delta q_1(t - i\Delta t_L, r) \tag{6}$$

, where, $N$ is the number of pulses. We used the following equation to calculate the temperature elevation:

$$q_N(t,\mathbf{r}) = \rho \int_{T_{ambient}}^{T_r} C_p(T) dT \tag{7}$$

, where, $\rho$ is the density (weight per unit volume), $C_p(T)$ is the specific heat under a constant pressure(J/g K), and $T_{ambient}$ is the ambient temperature (300 K). The room temperature value (5.4 × $10^{-3}$ cm$^2$/s) of the thermal diffusivity ($D_{th}$) of 60SiO$_2$-10Al$_2$O$_3$-30CaO(mol%) glass was used for all temperature ranges [43]. Although we assumed that energy loss due to thermal radiation can be negligible, we validated this assumption in the previous paper[42]. The density, $\rho$, of 60SiO$_2$-10Al$_2$O$_3$-30CaO(mol%). was 2.69 g/cm$^3$ [44]. The temperature dependence of $C_p$ for 62.5SiO$_2$-16.3Al$_2$O$_3$-0.05Na$_2$O-0.42K$_2$O-0.40FeO-10.5CaO-9.9MgO(mol%) [45] was used because no experimental values for melts with the same composition have been reported thus far, to the best of our knowledge. The central temperature $T_{center}$ was defined as the temperature midway between the two points of the lowest and highest temperatures of a specific zone under a temperature gradient; the Soret effect is discussed based on $T_{center}$.

### C. Molecular dynamics simulation

Non-equilibrium molecular dynamics simulations(NEMD) for 63.3SiO$_2$-16.3Al$_2$O$_3$-20.4CaO(mol%) glass were performed at $T_{center}$ = 2400 K and 4000 K using the LAMMPS package [46]. The simulation procedure was similar to that used in previous studies [25,26]. The ion numbers of the elements were 9795, 3165, 1020, and 1630 for O, Si, Al, and Ca, respectively, and the total number was 15610. The Coulomb-Buckingham-type two-body interatomic potential was used for the simulation:

$$V(r_{ij}) = \frac{1}{4\pi\varepsilon_0}\frac{z_i z_j e^2}{r_{ij}} + A_{ij} \exp\left(-\frac{r_{ij}}{\rho_{ij}}\right) - \frac{C_{ij}}{r_{ij}^6} \tag{8}$$

, where, $z$ is the partial charge, $\varepsilon_0$ is the electron permittivity constant of vacuum, $r$ is the distance between ion i and j. $A$, $B$, and $C$ are constant parameters that were determined by Du and Teter (TABLE SI) [47]. The first, second, and third terms represent the Coulombic interaction, short-range repulsive, and short-range attractive forces, respectively. The Coulombic term was calculated using the particle-particle-particle-mesh Ewald method. A periodic boundary condition was applied in the x, y, and z directions [48]. The time step was 1.0 fs. A cubic simulation box was used, and the cutoff for the short-range force was 12 Å. The simulation box was separated into eight slices, setting the fifth slice as the hotter slice and the first slice as the colder slice. First, the ions were positioned randomly and equilibrated at 4000 K using the velocity-rescaling method under a fixed volume[48]. Subsequently, the system was cooled to 2400 K using an NPT ensemble at a cooling rate of 6 K/ps under 500 MPa. The pressure was set to 500 MPa because the temperature elevation in the experiment occurred only around the focal spot inside the glass and the heated region was surrounded by rigid glass below the glass transition temperature [27]. Thereafter, we maintained the first slice at 2200 K and the fifth slice at 2600 K using the velocity-rescaling method with a constant volume [49]. At 4000 K, we eliminated the cooling process and maintained the first slice at 3800 K and the fifth slice at 4200 K.

Lesher [21] and Guy [50] examined the time evolution of the Soret coefficients of silicate melts. The following equation was used to explain the time dependence of the Soret coefficient of the $SiO_2$ component in the silicate melts:

$$\sigma_{\text{Soret},1}(t) = \sigma_{\text{Soret},1}^{\text{steady-state}} \left(1 - e^{-\frac{t}{\theta_{63\%}}}\right) \qquad (9)$$

, where $\sigma_{\text{Soret},1}^{\text{steady-state}}$ is the Soret coefficient at the steady state, and $\theta_{63\%}$ $(= L/(\pi^2 D_{\text{Si}}))$ is the characteristic time required for the Soret coefficient to be 63% of the steady state. $L$ and $D_{\text{Si}}$ are the distances between the colder and hotter regions and the self-diffusion coefficient of the Si ion, respectively. Si ions have the lowest diffusion coefficient among the ions in the silicate melt used in this study, and diffusion is the rate-determining factor. The calculated $D_{\text{Si}}$ values are shown in Fig. S2. The simulation time required for the system to reach steady state can be calculated using Eq.(9) and the mole-fraction sampling began at $\pi\theta_{63\%}$, which corresponds to 95% of the steady state, after the temperature gradient was applied. $T_{\text{center}}$ = 2400 K was used as the lowest temperature in the MD simulation because an excessive amount of calculation time was required for the system to accomplish a steady state below $T_{\text{center}}$ = 2400 K.

## III. RESULTS AND DISCUSSIONS

### A. Migration direction of SiO$_2$ component

The relative intensities of the X-rays obtained by EPMA are shown in Fig. 1. The laser propagation direction **k** is perpendicular to the images. The temperature around the central region is higher than that in the surrounding region [27,41,42]. For 250000 pulses (1 s) of irradiation, Si is concentrated in the central region, and Al and Ca are concentrated in the surrounding region. For the 800 pulse (3.2 ms) irradiation, Si is concentrated in the surrounding region, and Ca is concentrated in the central region; however Ca displays a small peak in the surrounding area. Al displays a peak in the surrounding area and a small peak in the central region. Qualitatively, the central region has a higher temperature than that of the outside region [27,41,42]; therefore, the above results indicate that $SiO_2$ easily migrates to the colder region when the irradiation time is 3.2 ms. An example of the vertical cross-sectional view of the element distribution is shown in Fig. S1 as a reference.

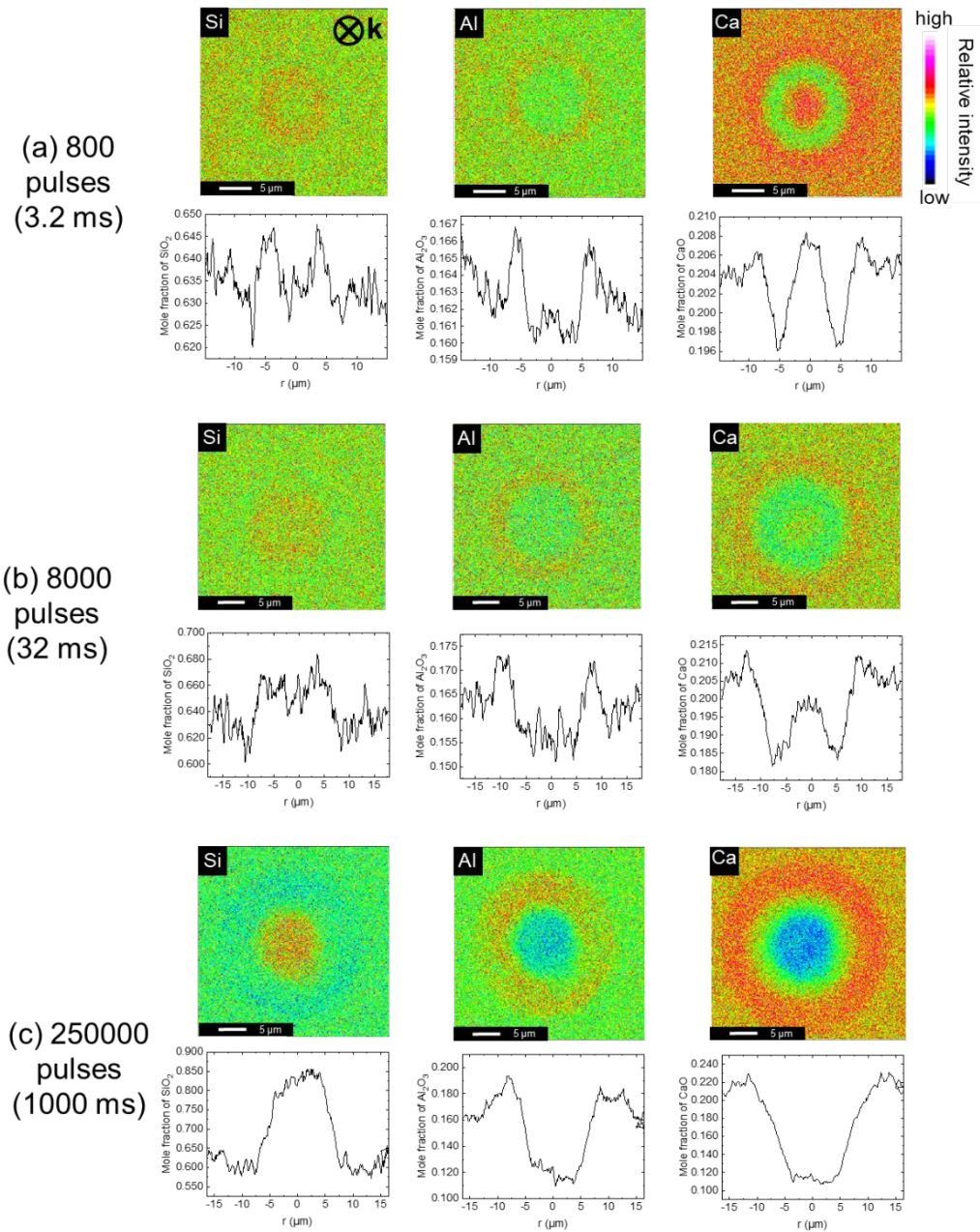

Fig. 1. Mole-fraction distribution of irradiated area measured by electron-probe micro analyzer. (a) 800 pulses, (b) 8000 pulses, and (c) 250000 pulses. The color images indicate the relative intensity of the X-ray signal. The line graph is the calculated concentration as the oxide component.

Generally, the temperature distribution under high-repetition femtosecond laser irradiation inside glass evolves with the irradiation time [27,41,42]. Repeating irradiation experiments were performed to eliminate the possibility that the temperature evolution from 800 pulses to 250000 pulses affects the migration direction of $SiO_2$ (Fig. 2). Upon increasing the repetition from one to thirty times, the element distribution qualitatively changes from 800 pulse like one to 250000 pulse like one (Fig. 1).

This occurrence indicates that the total irradiation time, rather than the evolution of the temperature distribution due to contentious irradiation, is important for determining the migration direction of SiO$_2$.

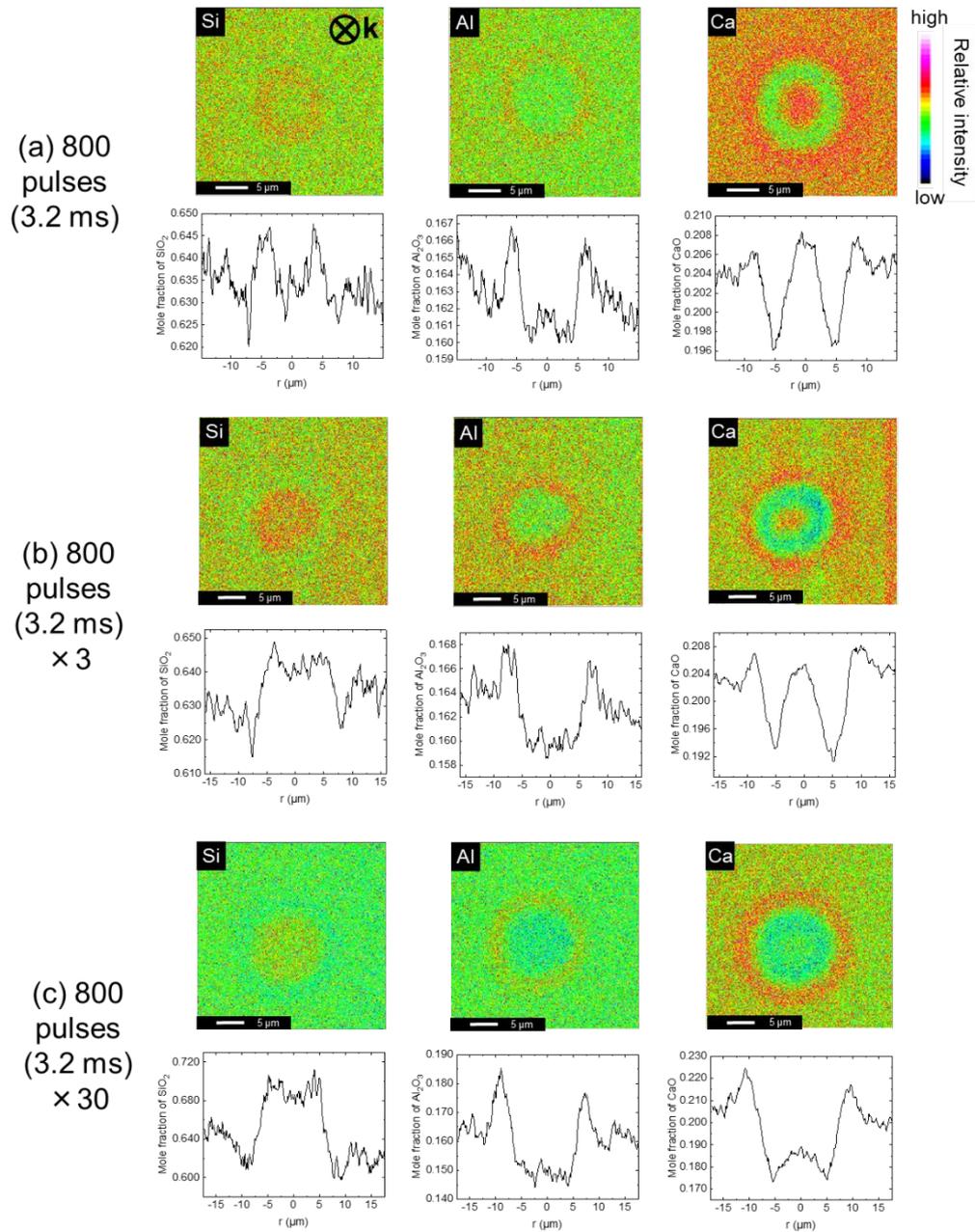

Fig. 2. Mole-fraction distribution of irradiated area measured by electron-probe micro analyzer. (a) 800 pulses×1, (b) 800 pulses×3, and (c) 800 pulses×30. The color images indicate the relative intensity of the X-ray signal. The line graph is the calculated concentration as the oxide component.

## B. Degree of progression of Soret effect in each temperature zone

For a more quantitative understanding, the degree of the progress of the Soret effect($\phi$) was defined as follows:

$$\phi = 1 - e^{-\frac{t}{\theta_{63\%}}} \tag{10}$$

Eq. (10) is a part of Eq. (9) and expresses a ratio of the Soret coefficient at a specific time to that at the steady state. $t$ was used as the laser irradiation time.

Fig. 3(a)–(c) show the relationships between the temperature distribution, migration zone, and concentration distribution of $SiO_2$. The peak $SiO_2$ concentration corresponds to 3060 K, which is the critical temperature($T_c$) at the boundary of the migration direction change. After 800 pulses (3.2 ms) irradiation, high-temperature zone (A) reaches $\phi = 100\%$, whereas medium-temperature zone (B) reaches only $\phi = 2.7\%$. This occurrence implies that the Soret effect occurs only in the high-temperature zone (A) under these conditions. However, after 250000 pulses (1000 ms) irradiation, medium-temperature zone (D) (below $T_c$) and high-temperature zone (C) reach $\phi = 100\%$ of the degree of progression. A summary of the degree of progression is presented in TABLE I. Note that a fixed temperature soon after the laser irradiation was used and the temperature evolution for calculating $\phi$ was neglected.

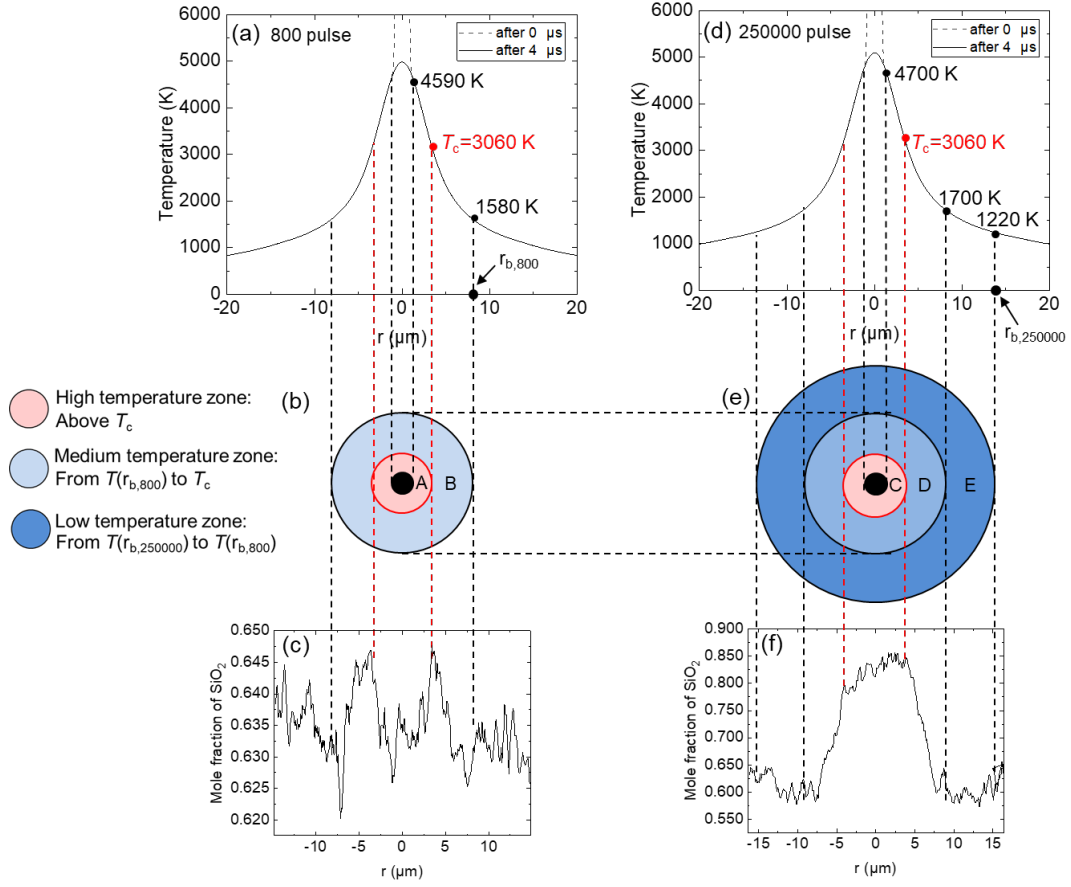

Fig. 3. Correspondence between temperature and composition after irradiation. (a), (d) Temperature distribution, (b), (e) Explanation and temperature of each zone, (c), (f) Mole-fraction of $SiO_2$ which is same as that in Fig. 1.

TABLE I. Degree of progression of Soret effect ($\phi$) in each zone and parameters for calculation.

| Zone | $t_{irrad}$ (ms) | $r_{min}$ (μm) | $r_{max}$ (μm) | $|r_{max}-r_{min}|$ (μm) | $r_{center}$ (μm) | $T$ at $r_{center}$ (K) | $D_{Si}$ (cm$^2$/s) | $\theta_{63\%}$ (s) | $\Phi$ (%) |
|---|---|---|---|---|---|---|---|---|---|
| A | 3.2 | 1.30 | 3.55 | 2.25 | 2.43 | 3830 | 3.70×10$^{-5}$ | 0.0001 | 100.00 |
| B | 3.2 | 3.55 | 8.25 | 4.70 | 5.90 | 2040 | 1.96×10$^{-7}$ | 0.1141 | 2.77 |
| C | 1000 | 1.30 | 3.74 | 2.44 | 2.52 | 3880 | 3.95×10$^{-5}$ | 0.0002 | 100.00 |
| D | 1000 | 3.74 | 8.25 | 4.51 | 6.00 | 2130 | 3.25×10$^{-7}$ | 0.0633 | 100.00 |
| E | 1000 | 8.25 | 14.10 | 5.85 | 11.18 | 1400 | 8.18×10$^{-10}$ | 42.3826 | 2.33 |

$t_{irrad}$, $r_{min}$, and $r_{max}$ are the irradiation time, minimum radius of the zone, and maximum radius of the zone, respectively. $r_{center}$ is defined as $(r_{min}+ r_{max})/2$. $D_{Si}$ is the self-diffusion coefficient of the Si ions calculated using MD shown in Fig. S2

Figure 4 shows a schematic summary explaining why short-time irradiation (800 pulses) leads to the migration of $SiO_2$ to the colder region, and long-time irradiation results in the migration of $SiO_2$ to the hotter region. Under short-time irradiation, the Soret effect occurs only in the high-temperature central zone above $T_c$ and Si diffuses to the colder region and stops. However, under long-time irradiation (250000 pulses), the same Soret effect occurs in the high-temperature zone during the early stage of irradiation. Thereafter, diffusion also occurs in the medium-temperature zone below $T_c$, and many Si ions migrate to the hotter region and are concentrated in that area. The temperature gradient is spherical; therefore, the medium-temperature zone has a larger volume and more atoms than that in the high-temperature zone. Thus, diffusion behavior in the medium-temperature zone below $T_c$ will be dominant during the late stage of irradiation.

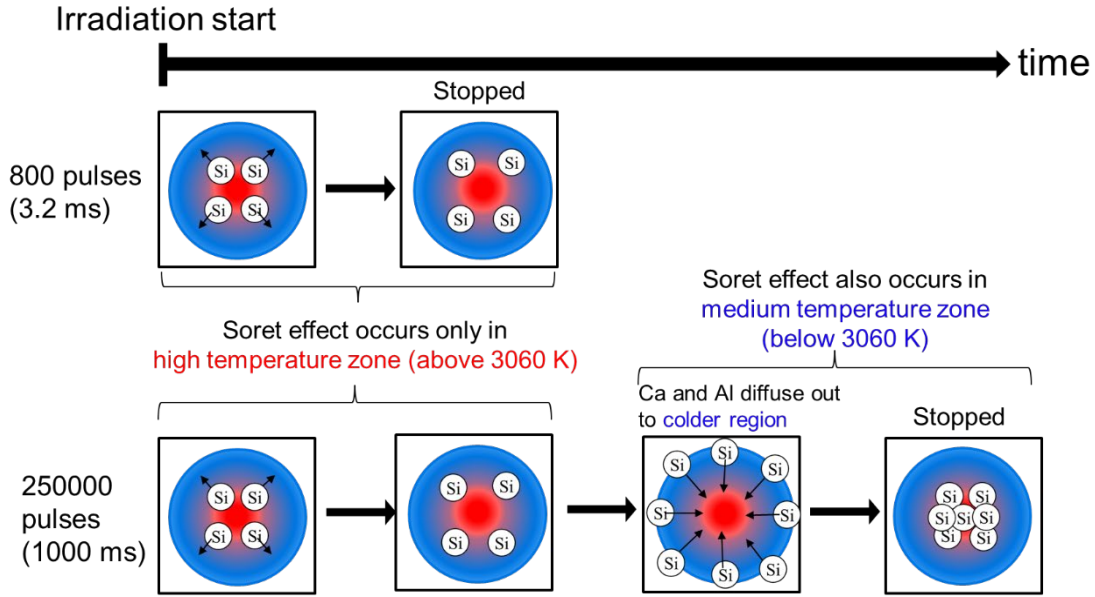

Fig. 4. Schematic illustration for mechanism of the Soret effect in this study.

**C. Molecular dynamics simulation**

NEMD simulations were performed to substantiate the experimental conclusion that $SiO_2$ migrates to the colder region of the high-temperature zone above $T_c$. The results for $T_{center}$ = 4000 K (Fig. 5(a)) indicate the migration direction of $SiO_2$ to the colder region, and correspond to the experimental result obtained above $T_c$ = 3060 K (Fig. 3(c)). The results for $T_{center}$ = 2400 K (Fig. 5(b)) indicate the migration direction of $SiO_2$ to the hotter region, and correspond to the experimental result obtained below $T_c$ = 3060 K (Fig. 3(c) and (f)).

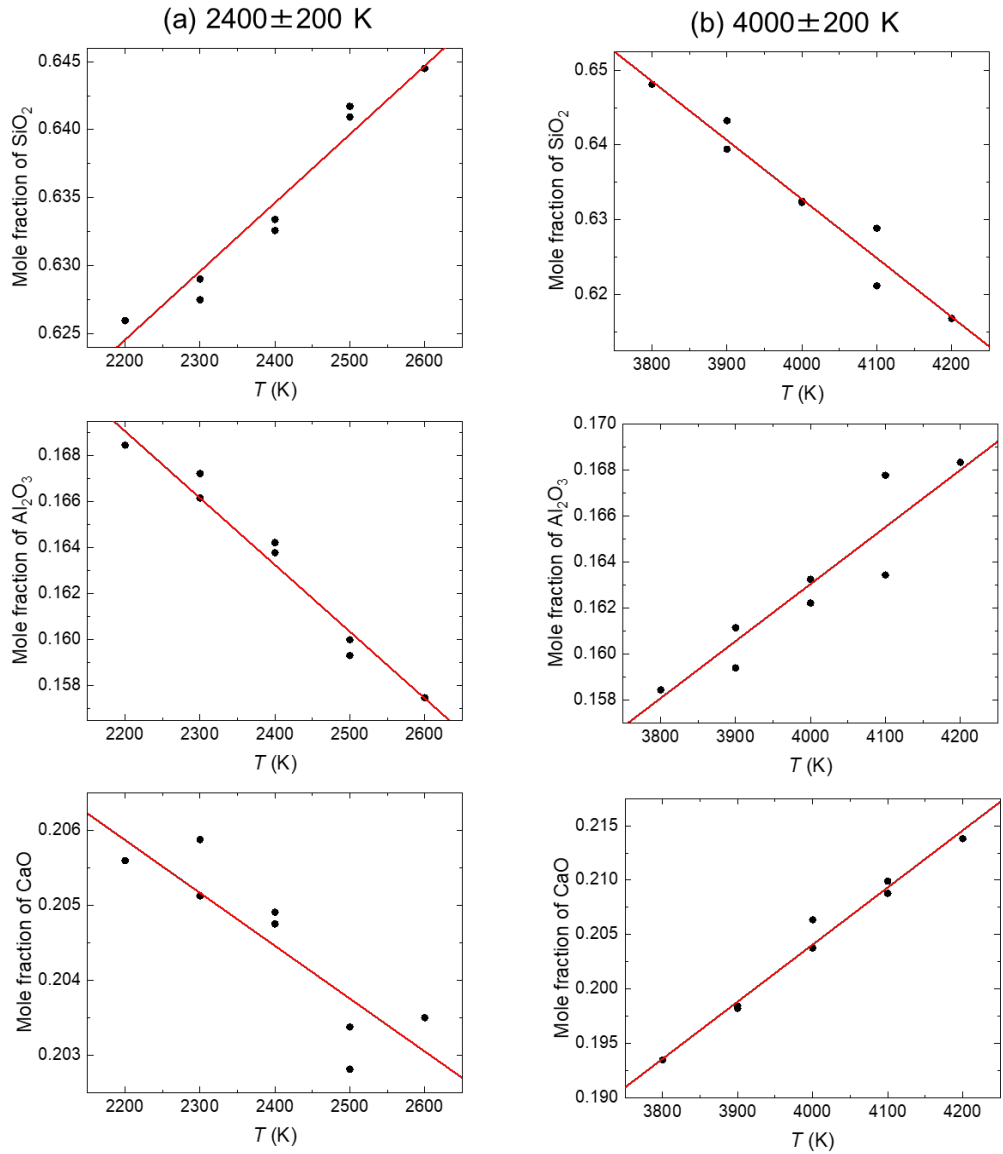

Fig. 5. Mole-fraction of the oxide of each element versus temperature as calculated by molecular dynamics simulation under a temperature gradient. (a) $T_{\text{center}}$ = 2400 K. (b) $T_{\text{center}}$ = 4000 K. The red lines are least-squares fits to the data.

Finally, we consider the reason for silica migration to the colder region at ultra-high temperatures above approximately 3000 K. Figure 5 shows the thermal expansion as calculated by molecular dynamics simulation without a temperature gradient; a bend in volume occurs in the range from 2000

K to 3400 K under 500 MPa, indicating that the second-order phase transition occurs within this temperature range. This result is consistent with the range of silica boiling point temperatures [51] (2503 K, 2863 K, and 3223 K), and indicates that the Soret effect in the non-liquid phase occurs at 4000 K. The self-diffusion coefficient and Voronoi volume of each element do not show marked changes with temperature (Fig. S2 and Fig. S3), indicating that the volatilization of specific elements does not occur. The partial pair distribution function does not show a notable change with temperature (Fig. S4), indicating that the static structure does not change markedly. Although the detailed mechanism has not been elucidated, we are inclined to believe that the kinetic mechanism will be different between 2400 K and 4000 K; additionally, the Si-O bond cutting procedure at 4000 K will be different from that at 2400 K. That is, the Si-O bond may be directly cut instead of the $S_N2$-like substitutional reaction [52].

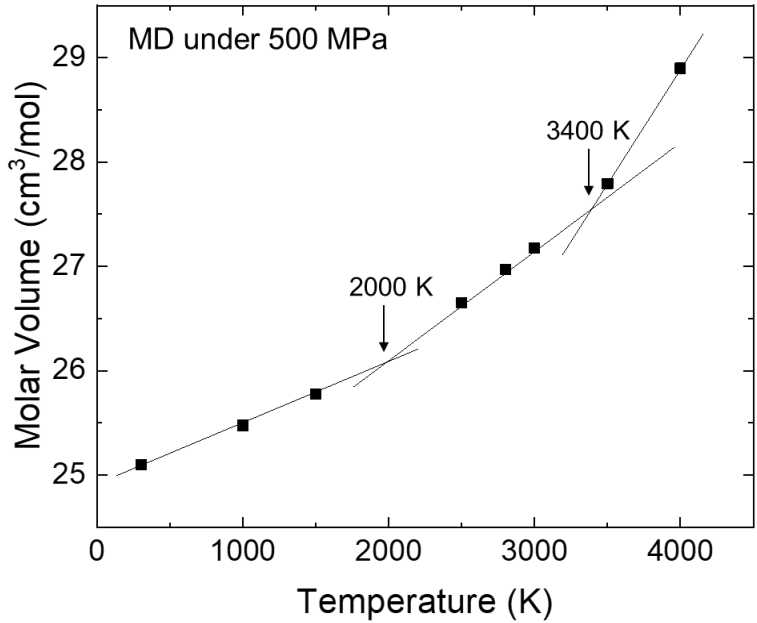

Fig. 5. Molar volume versus temperature as calculated by molecular dynamics simulation without a temperature gradient. The calculation was performed at 500 MPa.

## VI. CONCLUSION

In this study, we demonstrated that $SiO_2$ was concentrated in the colder region of silica melts under a temperature gradient at ultra-high temperatures above approximately 3000 K. The NEMD simulation showed that at $T_{center}$ = 4000 K, $SiO_2$ migrated to the colder region. The MD simulation showed that the second-order phase transition occurred at approximately 2000–3400 K, and would affect $SiO_2$ migration to the colder region. However, the detailed mechanisms remain to be elucidated.

**SUPPLEMENTARY MATERIAL**

See the supplementary material for the vertical cross-sectional view of the element distribution, potential parameters for molecular dynamics simulation, self-diffusion coefficient, Voronoi volume, partial pair distribution function, and running coordination number.

**ACKNOWLEDGMENTS**

This research was financially supported by the JSPS Foundation (20K05089).

**AUTHOR DECLARATIONS**

**Conflict of Interest**

The authors declare that there is no conflict of interest.

**DATA AVAILABILITY**

The data that supports the findings of this study are available within the supplementary material.

**Supporting Information**

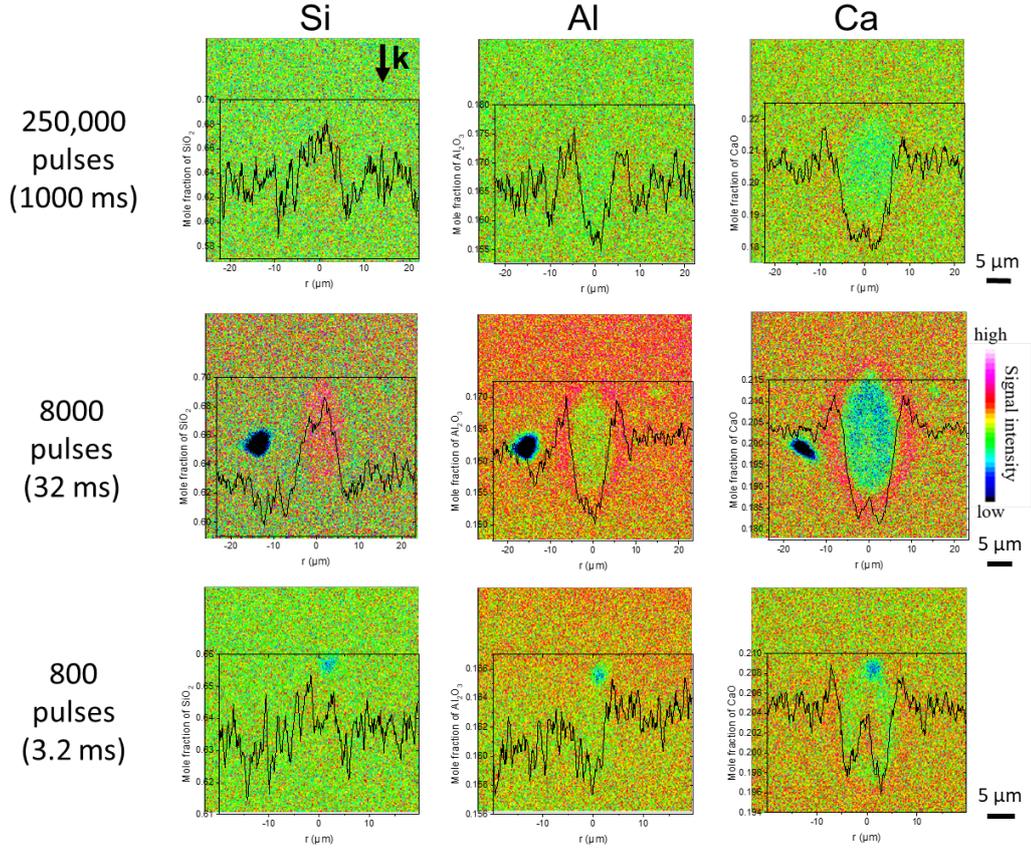

Fig. S1. Self-diffusion coefficient of each element in side view. The arrow **k** indicates the laser propagation direction.

TABLE SI. Potential parameter set used in this study as developed by Teter and Du [47].

| i–j | $A_{ij}$ (eV) | $\rho_{ij}$ (Å) | $C_{ij}$ (eV Å$^6$) |
| --- | --- | --- | --- |
| $O^{-1.2}$–$O^{-1.2}$ | 2029.2204 | 0.343645 | 192.58 |
| $Si^{+2.4}$–$O^{-1.2}$ | 13702.905 | 0.193817 | 54.681 |
| $Al^{+1.8}$–$O^{-1.2}$ | 12201.417 | 0.195628 | 31.997 |
| $Ca^{+1.2}$–$O^{-1.2}$ | 7747.1834 | 0.252623 | 93.109 |

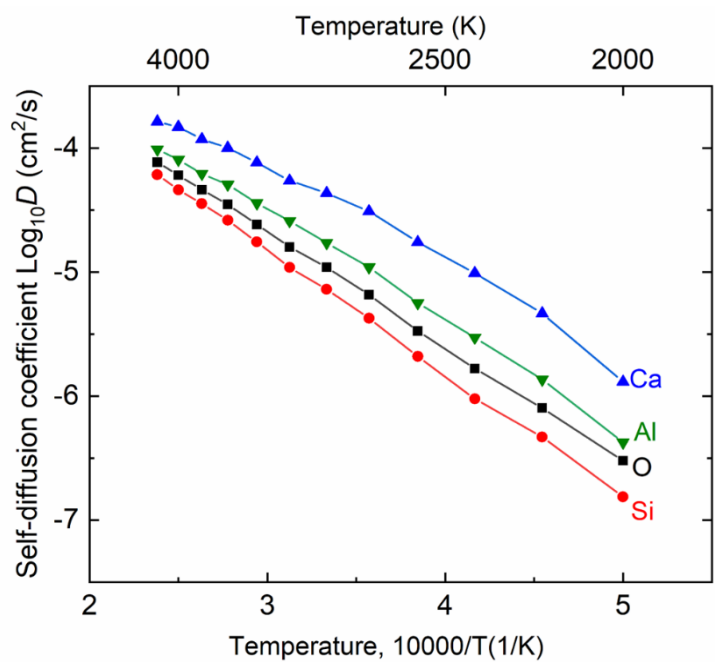

Fig. S2. Self-diffusion coefficient of each ion. The calculation was conducted under 500 MPa.

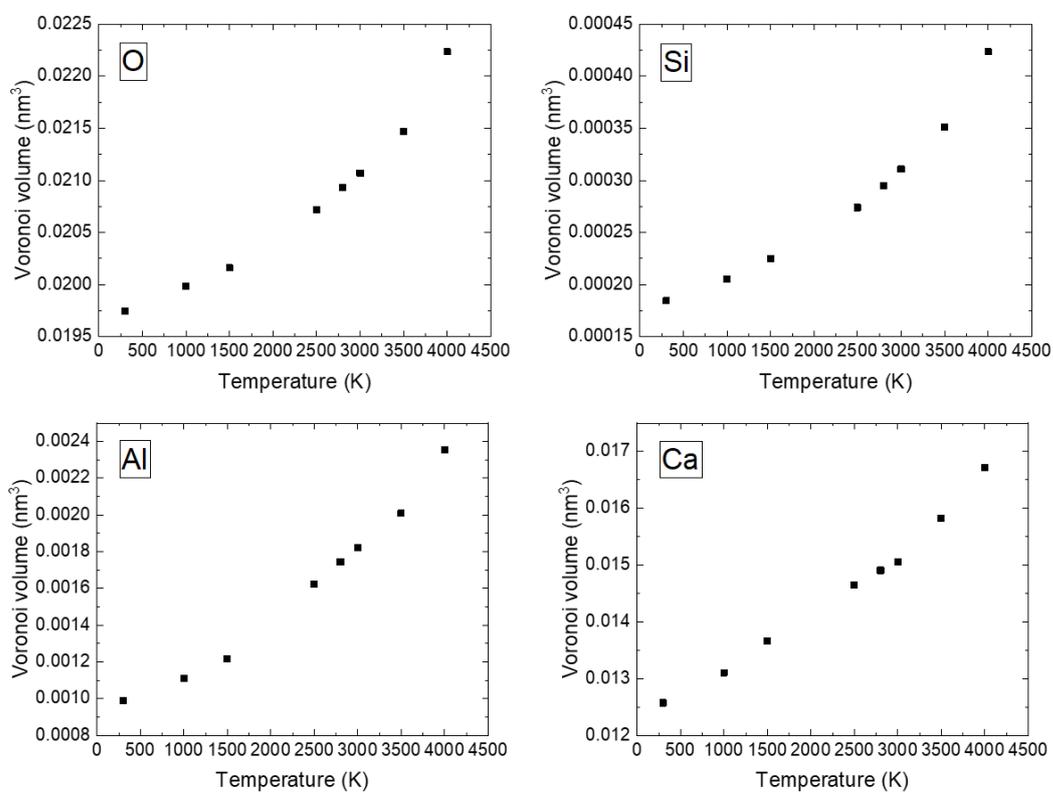

Fig. S3. Temperature dependence of Voronoi volume. The calculation was performed at 500 MPa, with a total ion number of 6244.

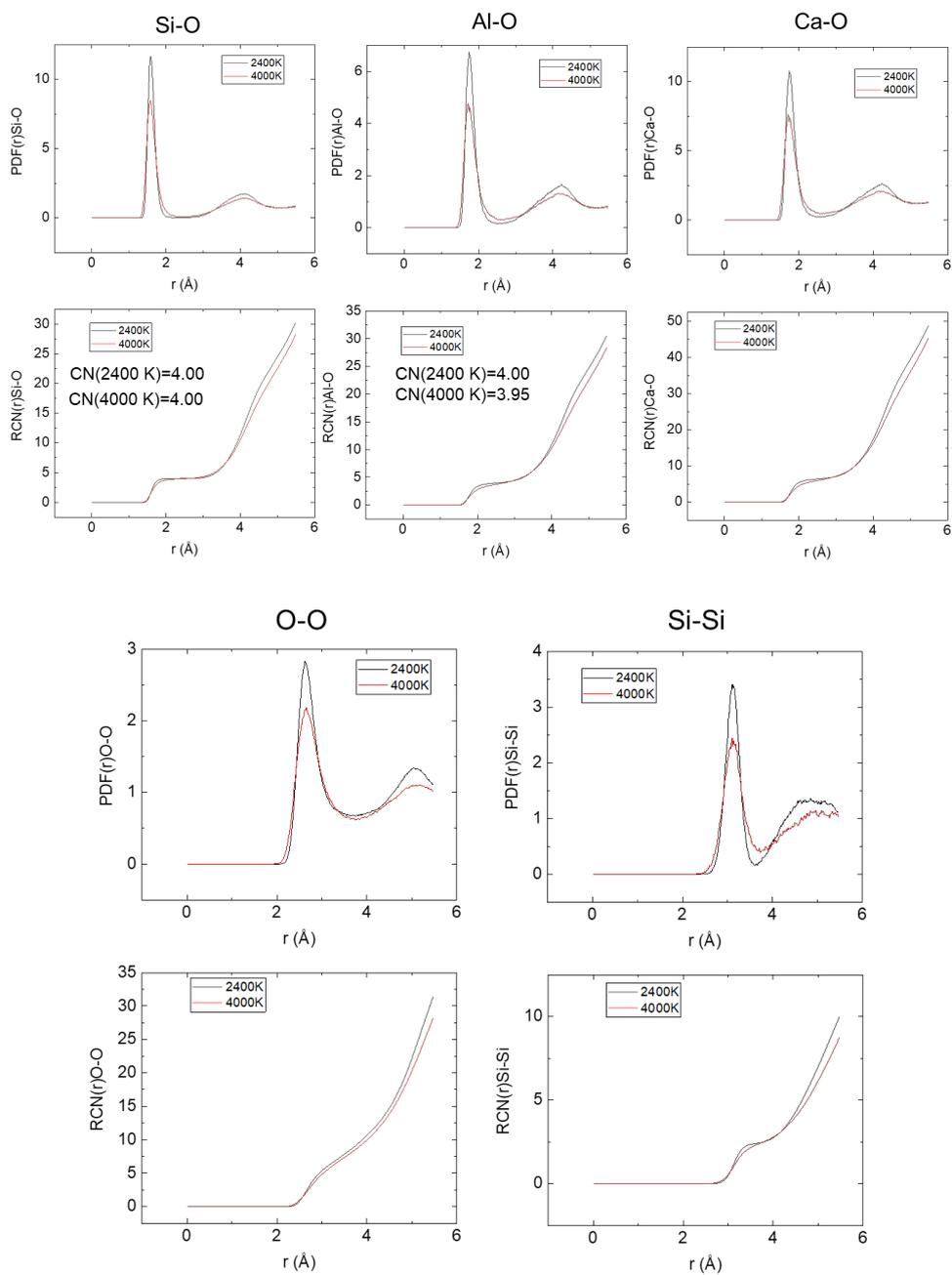

Fig. S4. Partial pair distribution function and running coordination number. The calculation was performed at 500 MPa, with a total ion number of 6244.